\documentclass{ws-mpla}
\usepackage[super]{cite}
\usepackage{graphicx}
\usepackage{amsmath}
\usepackage{amssymb}
\usepackage{float}
\usepackage{color}
\usepackage{dcolumn}
\usepackage{bm}
\usepackage{xcolor}
\usepackage{multirow}
\usepackage{adjustbox}
\usepackage{longtable}
\usepackage{comment}
\usepackage[normalem]{ulem}
\usepackage{array}
\usepackage[colorlinks,citecolor=blue,linkcolor=red,anchorcolor=blue,filecolor=blue,urlcolor=blue]{hyperref}
\usepackage{amssymb}
\usepackage{graphicx}
\usepackage{dcolumn}
\usepackage{bm}
\usepackage{caption}
\usepackage{subcaption}
\usepackage{subfloat}
\usepackage{float}
\usepackage{comment}
\usepackage{array}
\usepackage{color}
\usepackage{amsmath}
\usepackage[colorlinks,citecolor=blue,linkcolor=red,anchorcolor=blue,filecolor=blue,urlcolor=blue]{hyperref}
\begin{document}
\markboth{Praveen K. Yadav, Raj Kumar and M. Bhuyan}{Isospin properties of neutron-rich Sc, Ti, V, and Cr nuclei}
\catchline{}{}{}{}{}
\title{
Persistence of the N = 50 shell closure over the isotopic chains of Sc, Ti, V and Cr nuclei using relativistic energy density functional}
\author{Praveen K. Yadav, Raj Kumar}
\address{School of Physics and Materials Science, Thapar Institute of Engineering and Technology, Patiala, Punjab - 147004, India \\
praveenkumarneer@gmail.com and rajkumar@thapar.edu }

\author{M. Bhuyan}
\address{Center for Theoretical and Computational Physics, Department of Physics, Faculty of Science, University of Malaya, Kuala Lumpur - 50603, Malaysia \\ bunuphy@um.edu.my}

\maketitle


\begin{abstract}
The analytical expression of the density-dependent binding energy per nucleon for the relativistic mean field (RMF), the so-called relativistic energy density functional (Relativistic-EDF), is used to obtain the isospin-dependent symmetry energy and its components for the isotopic chain of $Sc$-, $Ti$-, $V$- and $Cr$-nuclei. The procedure of the coherent density fluctuation model is employed to formulate the Relativistic-EDF and Br\"{u}eckner energy density functional (Br\"{u}eckner-EDF) at local density. A few signatures of shell and/or sub-shell closure are observed in the symmetry energy and its components, i.e., surface and volume symmetry energy, far from the $\beta-$stable region for \textit{odd-A} $Sc$- and $V$-, and \textit{even-even} $Ti$- and $Cr$- nuclei with non-linear NL3 and G3 parameter sets. A comparison is made with the results obtained from Relativistic-EDF and Br\"{u}eckner-EDF with both NL3 and G3 for the considered isotopic chains. We find Relativistic-EDF outperforms the Br\"{u}eckner-EDF in predicting the shell and/or sub-shell closure of neutron-rich isotopes at N = 50 for these atomic nuclei. Moreover, a relative comparison has been made for the results obtained with the non-linear NL3 and G3 parameter sets.
\keywords{Relativistic Energy Density functional; Coherent Density Fluctuation Model; Isospin Asymmetry; Symmetry Energy}
\end{abstract}


\noindent
\section{Introduction} \label{intro}
The study of nuclei lying far away from the line of $\beta$-stability is of vital importance in the study of modern nuclear structure. Moving across the Seagr\`{e}  chart away from the line of stability, the limit of nuclear existence is reached. This limit represents the edges of the nuclear landscape where nuclear binding ends and the nuclei are unbound and disintegrate quasi-instantaneously to neutron or proton emission in the ground state, which corresponds to the neutron and proton drip-line, respectively \cite{afan15,neuf19,erle12}. Nuclei that lie far away from the $\beta$-stability line are highly asymmetric nuclei with large differences in the number density of neutrons and protons. Recent advances in radioactive ion beam facilities and highly sensitive detector mechanisms have accelerated the research of extreme $N/Z$ (asymmetric) nuclei, which helps to increase the odds of experimental understanding of drip-line nuclei in terms of isospin asymmetry, which is largely unexplored as yet \cite{muts17,bhuy12,ohni08,bhuy15,jens04,hans03,pfue12,gaff55}. Recently, the link between the degrees of freedom of the nuclear shape and the emergence of the neutron drip line \cite{tsun20}, which stimulates further understanding of nucleosynthesis in the case of neutron-rich nuclei, has been suggested. Furthermore, it is practically impossible to increase the number of neutrons indefinitely for any finite nucleus \cite{erle12}. This statement can be realized as we move towards the neutron drip-line on the Seagr\`{e} chart; an increase in the number of \textit{n-n} pairs is observed as compared to \textit{n-p} pairs. Consequently, following Pauli's principle, a large amount of repulsive energy leads to instability in the nuclides. In Ref. \cite{satp04}, the shell effects liberate the nucleus by overpowering the instability owed to the repulsive effect of nuclear force. These shell effects led to the discovery of new magic numbers, the island of inversion, and the associated stability near the drip-line.

The nuclear symmetry energy coefficient is one of the most important quantities for understanding the isospin-dependence in nuclear structure analysis for finite and infinite nuclear matter. The nuclear matter equation of state (EoS), which is directly related to the isospin asymmetry, is directly connected to the symmetry energy \cite{stei05,bhuy13,kuma18}, and can be defined as the amount of energy required to convert symmetric nuclear matter involving the same number of protons and neutrons into pure neutron matter. It plays a significant role in different areas of nuclear physics such as giant collective excitation \cite{rodi07,bisw15}, heavy-ion reaction dynamics \cite{chen08,colo09,li08}, the ground-state structure of nuclei \cite{Niki08,Giai10,dale10,sarr07}, and even the astrophysics of neutron star systems \cite{stei05,fatt12,dutr12,dutr14}. Over the years, significant work has been carried out to correlate the surface properties of nuclear matter with finite nuclei. This has largely helped to constrain the magnitude of the symmetry energy $S$ and the slope parameter $L_{sym}^{A}$ at the saturation density $\rho_{o}$ of nuclear matter \cite{bara05,li08,tsan12,trau12,Horo14}. Furthermore, the critical difference between finite nuclei and infinite nuclear matter is that the latter lacks the Coulomb interaction. It is possible to characterize the nuclear matter as a function of energy density per particle \cite{Giai10}. In the present scenario, studying nuclear symmetry energy is crucial to developing a robust understanding of nuclear properties, including stability close to the drip-line, which is largely unexplored. A wide range of research work is underway on the density dependence of the symmetry energy and its dependent quantities \cite{bhuy18,anto16,qudd20}. The study of symmetry energy of finite nuclear systems at local density is performed using the EoS of asymmetric nuclear matter by employing conventional methods, including the liquid drop model, the Skyrme energy density functional, and the relativistic mean-field using the random phase approximation, among others. Alternatively, the Br\"{u}eckner-EDF within the coherent density fluctuation model (CDFM) has been largely successful in studying the surface properties of nuclei in regions of light, heavy and superheavy mass \cite{gaid11,gaid12,anto16,bhuy18,qudd20}. Although the early idea of CDFM dates long back, the profound implications of the work still need to be realized.

In recent years, the surface properties of nuclei can be calculated by incorporating either the non-relativistic or relativistic inputs within the CDFM \cite{anto94, gaid11, gaid12, bhuy18, akum21}. These studies found significant indications for possible shell and/or sub-shell closure along the isotopic chains. Although CDFM with non-relativistic Br\"{u}eckner's prescription can predict a few shell/sub-shell closures at and/or near drip-line in terms of isospin-dependent quantities such as the symmetry energy and its coefficients. It fails to justify the traditional neutron magic number ($N$ = 126) for $^{208}$Pb \cite{jeet22}. By correlating the above facts, one can raise the issue that the classical Br\"{u}eckner's prescription fails to accurately replicate the empirical saturation point of symmetric nuclear matter, that is $E/A \approx$ -16 MeV at $\rho \approx 0.2$ fm$^{-3}$ instead of $\rho \approx 0.15$ fm$^{-3}$, commonly known as the Coester band problem \cite{coes70,brock90}. Thus, it is important to adopt a method that can help mitigate the Coester-band problem \cite{coes70,brock90}. Therefore, an alternative method is set up that involves the inclusion of a functional relativistic energy density functional by following the Br\"{u}eckner's approach. In previous works of one of us and collaborators, using a new fitting procedure for Relativistic-EDF \cite{akum21,jeet22}, the surface properties of some doubly magic nuclei and $Pb$- isotopic chain were explored. Recently, a great deal of research has been reported with nuclei close to the proton number $Z$ = 20 (Calcium), showing the possible emergence of new shell and/or sub-shell closure at $N$ = 32, 34 and 40 regions, which indicate the possible candidature for $Sc$ ($Z$ = 21) \cite{leis21}, $Ti$ ($Z$ = 22) \cite{jans02,mich20,leis18,rodr20}, $V$ ($Z$ = 23) \cite{reit18,napo06}, $Cr$ ($Z$ = 24) \cite{pris01,burg05}. In the present analysis, we estimate the density-dependent symmetry energy along with its surface and volume components for the $odd-A$ $Sc$- ($Z = $21) and $V$- ($Z = $23), and $even-even$ $Ti$ ($Z = $22) and $Cr$- ($Z = $24) isotopic chains by using the newly fitted Relativistic-EDF \cite{akum21} for the widely used non-linear NL3 \cite{lala97} and newly developed G3 \cite{bkum17,kuma18} parameter sets that replicates the experimental data consistently with comparatively minor fluctuations for a wide range of nuclei.

This article is organized as follows: In Section \ref{theory}, we present a brief description of the relativistic mean-field model along with the coherent density fluctuation model (CDFM). Section \ref{Results} discusses the calculations and results. Section \ref{Summary} contains a summary and conclusions.

\section{Theoretical Formalism} \label{theory}
\label{Theory} \noindent
This section describes, in brief, the relativistic mean-field (RMF) theory. The RMF model involves the use of a microscopic approach for solving the many-body problem via the interacting meson fields. The RMF Lagrangian density is constructed by the interactions of isoscalar-scalar $ \sigma $-, isoscalar-vector $\omega$-, isovector-scalar $\delta$-, and isovector-vector $\rho$-mesons with the nucleons along with the inclusion of crossed coupling of mesons up to the fourth order \cite{sero86,bogu77,bhuy09,bhuy15,bhuy18,lala97,lala99,akum21,kuma18,bkum17,furn97,furn96,broc92,fuch95,carl00}. Following the RMF energy density and applying the Euler-Lagrange equation, one can get the equations of motions for mesons and nucleons. A group of coupled differential equations is obtained that are solved self-consistently \cite{kuma18}. The scalar and vector densities can be written as:
\begin{eqnarray}
\centering 
\label{eqn:1}
\rho_{s}(r) & = &\sum_{\alpha} \varphi_{\alpha}^{\dagger}(r) \beta \varphi_{\alpha}, \\
\rho_{3}(r) & = &\sum_{\alpha} \varphi_{\alpha}^{\dagger}(r) \tau_{3} \varphi_{\alpha},
\end{eqnarray}
which are evaluated based on the converged solutions within the spherical harmonics. The vector density $\rho_{3}(r)$ is later used within the CDFM formalism to estimate the weight function $\vert \mathcal{F}(x) \vert^{2}$. The weight function is further used to estimate the density-dependent symmetry energy (\textit{S}) and its associated parameters. 

The energy density of infinite and isotropic NM is construed using the energy-momentum tensor as:
\begin{eqnarray}
	\centering 
	\label{eqn:3}
	T_{\mu \nu}=\sum_{i} \partial_{\nu} \phi_{i} \frac{\partial \mathcal{L}}{\partial\left(\partial^{\mu} \phi_{i}\right)}-g_{\mu \nu} \mathcal{L}.
\end{eqnarray}
The energy density of the system as a function of the scalar density $\rho_{s}(r)$ and the vector densities $\rho_{3}(r)$ is obtained by the zeroth component of the energy-momentum tensor $T_{00}$. The energy density functional for an interacting system of nucleon-meson $\mathcal{E}(k)_{\text {nucl.}}$ can be written as \cite{akum21}:
\begin{eqnarray}
	\begin{aligned}[b]
		\mathcal{E}(k)_{\text {nucl.}} = & \frac{2}{(2 \pi)^3} \int d^3 k E_i^*(k)+\frac{m_s^2 \Phi^2}{g_s^2}\left(\frac{1}{2}+\frac{\kappa_3}{3 !} \frac{\Phi}{M}+\frac{\kappa_4}{4 !} \frac{\Phi^2}{M^2}\right) \\
		&+\rho_b W-\frac{1}{4 !} \frac{\zeta_0 W^4}{g_\omega^2}-\frac{1}{2} m_\omega^2 \frac{W^2}{g_\omega^2}\left(1+\eta_1 \frac{\Phi}{M}+\frac{\eta_2}{2} \frac{\Phi^2}{M^2}\right) \\
		&+\frac{1}{2} \rho_3 R-\frac{1}{2}\left(1+\frac{\eta_\rho \Phi}{M}\right) \frac{m_\rho^2}{g_\rho^2} R^2-\Lambda_\omega\left(R^2 \times W^2\right) \\
		&+\frac{1}{2} \frac{m_\delta^2}{g_\delta^2} D^2 .
	\end{aligned}
	\label{eqn:4}
\end{eqnarray}
Here, the terms $\Phi$, $R$, $W$,  and $D$ are the fields for $\sigma$, $\rho$, $\omega$, and $\delta$ mesons stated as $\Phi=g_s \sigma_0$, $R=g_\rho \vec{\rho}_0{ }^\mu$, $W=g_\omega \omega_0$, and $D=g_\delta \delta_0$, respectively. Moreover, $M$, $m_\sigma$, $m_\rho$, $m_\omega$, and $m_\delta$ are the masses of nucleon, $\sigma$, $\rho$, $\omega$, and $\delta$ mesons, respectively.

This model is found to be quite successful in predicting the structural properties of the ground and the intrinsic excited state across the Seagr\`{e} chart. Moreover, in recent years, the application of RMF formalism has gained widespread recognition in the study of nuclear astrophysics. The NL3 force parameter \cite{lala97} is one of the oldest and still widely used relevant force parameters in the study of finite nuclei and infinite nuclear matter. It considers the self-coupling of the $\sigma$-meson. Recently, the G3 parameter \cite{bkum17} based on effective field theory motivated relativistic mean-field formalism was put forward, which included various possible self- and cross-coupling terms of different mesons and featured the inclusion of $\delta$-mesons. Including $\delta$-meson is crucial in the highly asymmetric nuclear matter such as neutron stars. However, it also gained prominence in studying finite nuclei in heavy and super-heavy regions. Following Ref. \cite{ bkum17,kuma18}, the root-mean-square (\textit{rms}) deviation for the binding energy corresponding to 70 spherical nuclei for NL3 and G3 parameter sets is 2.977 MeV and 2.308 MeV, respectively. Moreover, the \textit{rms} deviation of charge radii obtained for eight close shell nuclei is 0.021 fm and 0.018 fm for NL3 and G3 force parameters, respectively. More details concerning the RMF Lagrangian and their parametrization are provided in Refs. \cite{sero86,bogu77,bhuy09,bhuy15,bhuy18,lala97,lala99,akum21,kuma18,bkum17,furn97,furn96}.
 
\subsection{Br\"{u}eckner prescription and relativistic energy density functional} 
\label{sssec:num1} 
{In order to perform a detailed study on the surface properties of isotopic chains, we present a qualitative description of the Br\"{u}eckner's prescription and effective field theory motivated relativistic mean-field (E-RMF) model fitting procedure \cite{akum21} and its application within the coherent density fluctuation model \cite{anto79,anto82,gaid11} in the following subsections. An important aspect of the fitting procedure is to translate the nuclear matter quantities from momentum space to coordinate space and reconstruction of nuclear matter quantities at local density. The calculation assumes that the nuclear matter is composed of tiny spheres of nuclear matter, which are referred to as ``fluctons" with local density function $\rho_{o}(x)=\dfrac{3A}{4\pi x^{3}}$ \cite{gaid11, gaid12,anto79,anto82,bhuy18}. Within the Br\"{u}eckner's prescription, the expression pertaining to the energy density of infinite and isotropic nuclear matter is derived using \cite{brue68,brue69}
\begin{eqnarray}
	\centering 
	\label{eqn:Vx}
	V(x) =AV_{0}(x)+V_{C}-V_{CO},
\end{eqnarray}
where
\begin{eqnarray}
	\begin{aligned}[b]
		V_{0}(x)=& 37.53\left[(1+\delta)^{5 / 3}+(1-\delta)^{5 / 3}\right] \rho_{o}^{2 / 3}(x) \\
		&+b_{1} \rho_{o}(x)+b_{2} \rho_{o}^{4 / 3}(x)+b_{3} \rho_{o}^{5 / 3}(x) \\
		&+\delta^{2}\left[b_{4} \rho_{o}(x)+b_{5} \rho_{o}^{4 / 3}(x)+b_{6} \rho_{o}^{5 / 3}(x)\right],
	\end{aligned}
	\label{eqn:V0x}
\end{eqnarray}
with
\begin{equation}
	\begin{array}{lll}
		b_{1}=-741.28, & b_{2}=1179.89, & b_{3}=-467.54 \\
		b_{4}=148.26, & b_{5}=372.84, & b_{6}=-769.57.
	\end{array}
	\label{eqn:10}
\end{equation}
In Eq. \ref{eqn:Vx}, $ V_{0}(x)$ refers to energy per nucleon (MeV) corresponding to nuclear matter, which accounts for the neutron-proton asymmetry and the term $ V_{C}$ corresponds to the Coulomb energy of protons within a flucton given as:
\begin{equation}
	V_{C}=\frac{3}{5} \frac{Z^{2} e^{2}}{x},
	\label{eqn:11}
\end{equation}
and the Coulomb exchange energy is 
\begin{equation}
	V_{CO}=0.7386 Z e^{2}\left(3 Z / 4 \pi x^{3}\right)^{1 / 3}.
	\label{eqn:12}
\end{equation}
The crucial part of the calculation involves the translation of nuclear matter quantities of Eq. \ref{eqn:Vx} from momentum space to coordinate space within the local density approximation. In the present case, we are concerned with calculating the symmetry energy and its components. The most general form of nuclear symmetry energy ($S^{NM}$) can be expressed as,
\begin{eqnarray}
	\centering 
	\label{eqn:6}
	S^{NM}(\rho)=\dfrac{1}{2}\dfrac{\partial^{2}(\mathcal{E}/\rho)}{\partial\alpha^{2}}\bigg|_{\alpha=0}.
\end{eqnarray}
Following Refs. \cite{gaid11, gaid12}, the nuclear symmetry energy at local density can be obtained by using the Eqs. \ref{eqn:V0x} and \ref{eqn:6} as:
\begin{eqnarray}
	S^{\mathrm{NM}}(x)= 41.7 \rho_{o}^{2 / 3}(x)+b_{4} \rho_{o}(x)+b_{5} \rho_{o}^{4 / 3}(x)+b_{6} \rho_{o}^{5 / 3}(x).
	\label{eqn:13}
\end{eqnarray}
A detailed approach regarding the Br\"{u}eckner prescription of energy density functional which we have used for comparison, can be found in Refs. \cite{brue68,brue69,gaid11,gaid12}.

By adopting the CDFM approach, the fitted binding energy function at local density within the RMF formalism can be expressed as:
\begin{eqnarray}
	\centering 
	\label{eqn:5}
	\mathcal{E}(x)=C_{k} \rho_{o}^{2 / 3}(x)+\sum_{i=3}^{14}\left(b_{i}+a_{i} \alpha^{2}\right) \rho_{o}^{i/3}(x).
\end{eqnarray}
Here $C_{k}$ represents the coefficient of kinetic energy following the Thomas-Fermi approach and is given as $C_{k} = 37.53[(1+\alpha)^{5/3}+(1-\alpha)^{5/3}]$. The term $\alpha$ represents asymmetry in baryon density which is given as $\alpha = \left (\frac{\rho_n - \rho_p}{\rho_n + \rho_p} \right)$, where $\rho_n$ and $\rho_p$ are the densities of neutron and proton respectively \cite{stei05}. For finding the exact nature of binding energy per particle \textit{(E/A)} in position space, polynomial fitting with several terms is used (Eq. \ref{eqn:5}). The mean deviation is calculated from the expression, $\delta = [\sum_{j=1}^{N}(E/A)_{j,Fitted}-(E/A)_{j,RMF}]/N$. Here, $(E/A)_{j,Fitted}$ refers to the binding energy inferred from Eq. \ref{eqn:5}, $(E/A)_{j,RMF}$ refers to the binding energy per particles inferred from RMF functional and $n$ refers to the number of data points. Based on the analysis in Ref. \cite{akum21}, it is found that the use of 12 terms obtains the best fit. The coefficients inferred from the polynomial fitting is given in Table-I of the Ref. \cite{akum21}. Using Eqs. \ref{eqn:6} and \ref{eqn:5}, 
\begin{eqnarray}
	\centering 
	\label{eqn:7}
	S^{NM}(\rho)=41.7 \rho_{o}^{2/3}(x)+\sum_{i=3}^{14}a_{i}\rho_{o}^{i/3}.
\end{eqnarray}	
The above-defined quantities in Eqs. (\ref{eqn:6})-(\ref{eqn:7}) are obtained by following the procedure adopted for Br\"{u}eckner's functional as given in the Refs. \cite{brue68,brue69}. The densities of $Sc$-, $Ti$-, $V$- and $Cr$- nuclei are calculated using the RMF formalism for NL3 and G3 parameter sets and subsequently utilized as input in the CDFM for the calculation of the weight function. The weight function serves as a crucial quantity that bridges nuclear matter parameters existing in the momentum space and the coordinate space. In order to translate the momentum space to the coordinate space, one can make use of the CDFM formalism as discussed in the following subsection.

\subsection{Coherent density fluctuation model} 
\label{sssec:num2} 
The coherent density fluctuation model (CDFM) formulated by Antonov {\it et al.} \cite{anto79,anto82} is a natural extension of the Fermi-gas model which accounts for the fluctuations of momentum and coordinate space via the uncertainty principle. It is a well-established formalism and is based upon the $ \delta $-function limit of the generator coordinate method \cite{grif57}. This model can easily provide symmetry energy along with its surface and volume derivatives for finite nuclei. 
By weighting the corresponding quantities of infinite nuclear matter in momentum space within the CDFM, the symmetry energy coefficient for a finite nucleus in coordinate space can be written as \cite{anto94, gaid11, gaid12, bhuy18}:
\begin{eqnarray}
	\label{eqn:14_cdfm}
	S = \int_{0}^{\infty}dx \vert\mathcal{F}(x)\vert^{2}S^{NM}[f_{\rho}(x)].
\end{eqnarray}
The term $\vert\mathcal{F}(x)\vert^2$ in Eq. \ref{eqn:14_cdfm} refers to a weight function which is given by the expression
\begin{eqnarray}
	\centering 
	\label{eqn:15_cdfm}
	\vert \mathcal{F}(x) \vert^{2} = -\bigg(\frac{1}{f_{\rho}(x)}\frac{d\rho(r)}{dr} \bigg)_{r=x},
\end{eqnarray}
with the normalization as $\int_{0}^{\infty}dx \vert \mathcal{F}(x) \vert^{2} = 1$ and $S^{NM}[f_{\rho}(x)]$ refers to the symmetry energy of
nuclear matter with density $f_{\rho}(x)$ \cite{anto94,gaid11, gaid12, bhuy18}. The term $f_{\rho}(x)$ refers to the spherical radius generator coordinate for all $ A $ nucleons contained within the uniformly distributed spherical Fermi gas given as  $f_{\rho}(x)=\dfrac{3A}{4\pi x^{3}} $. The detailed analytical derivation of density-dependent symmetry energy using CDFM can be followed in Refs. \cite{anto79,anto82,anto94,bhuy18}.
Further, within Bethe-Weizs\~{a}cker liquid drop model, the symmetry energy can be expressed in terms of volume and surface symmetry energy as \cite{dani09},
\begin{eqnarray}
	\centering 
	\label{eqn:16_cdfm}
	S = \dfrac{S_{V}}{1+\dfrac{S_{S}}{S_{V}}A^{-1/3}}=\dfrac{S_{V}}{1+\dfrac{1}{\kappa A^{1/3}}}. 
\end{eqnarray}
The term $\kappa$ is the ratio of volume symmetry energy to that of the surface symmetry energy. From the Eq. \ref{eqn:16_cdfm} the individual components of symmetry energy, namely volume, and surface symmetry energy, can be written as:
\begin{eqnarray}
	\centering 
	\label{eqn:17_cdfm}
	S_{V}=S\bigg(1+\dfrac{1}{\kappa A^{1/3}}\bigg)
\end{eqnarray}
and    
\begin{eqnarray}
	\centering 
	\label{eqn:18_cdfm}
	S_{S}=\dfrac{S}{\kappa}\bigg(1+\dfrac{1}{\kappa A^{1/3}}\bigg),
\end{eqnarray}
respectively. Here we select to use Danielewicz's prescription \cite{dani03, dani06} over the liquid drop model (LDM) in order to split the volume and surface components of the symmetry energy. The symmetry energy coefficients have been precisely constrained when applying a combination of data available on binding energies, neutron skin thickness, and isobaric analogue states of finite nuclei from the mass formula by Danielewicz \cite{dani03, dani06}. The term $\kappa$ is calculated using the expression stated as \cite{anto16,danc20}, 
\begin{eqnarray}
	\centering 
	\label{eqn:19_cdfm}
	\kappa=\dfrac{3}{r_{0} \rho_{0}}\int_{0}^{\infty}dx \vert\mathcal{F}(x) \vert^{2} x f_{\rho}(x) \bigg[\dfrac{S^{NM}(\rho_{0})}{S^{NM}[f_{\rho}(x)]}-1\bigg],
\end{eqnarray} 
In Eq. \ref{eqn:19_cdfm}, $r_{0}$ is the radius of the nuclear volume per nucleon, $ \rho_{0} $ refers to the nuclear matter equilibrium density \cite{diep07,anto16} and $S^{NM}(\rho_{0})$ refers to symmetry energy at equilibrium density \cite{anto16,danc20}.\\

Recently, an alternative method was proposed to calculate the volume and surface symmetry energy within the CDFM formalism by incorporating non-relativistic inputs in the weight function \cite{gaid21}. The main motivation of the alternate method discussed in Ref. \cite{gaid21} is to ensure the accurate behaviour of the terms in the denominator of Eq. \ref{eqn:16_cdfm} in the limit of infinite nuclear matter. In the alternate method, as the limit of $A$ approaches infinity, the ratio $S_S/S_V$ tends toward zero. Consequently, the term $\left[S_S/S_V\right] A^{-1/3}$ also tends towards zero, and the symmetry energy approaches the correct limit of $S$, tending towards $S_V$. Whereas, while using the present approach, we require a condition to be applied, namely, the surface coefficient $S_S(A)$ must decrease to zero at a slower rate than $A^{-1/3}$ as $A$ approaches infinity. The alternate approach replaces the density $\rho(r)$ for the half-infinite nuclear matter with the density of a finite nucleus. Moreover, the Eq. (\ref{eqn:16_cdfm}) shows singularity for some of the potentials. The choice of the integration interval, particularly the lower limit of integration, affects the sensitivity of the results for $\kappa$ as in Eq. (\ref{eqn:16_cdfm}). The mathematical ambiguity in using the present approach can be relevant when dealing with exceedingly high mass numbers. The alternate method aims to provide a larger value of surface symmetry energy while providing a smaller value of volume symmetry energy. To note, the effect of using the alternate formalism may result in a marginally better-constrained range of surface and volume terms for some nuclei. Nonetheless, both formalisms provide identical evidence of shell and/or sub-shell closure in the finite nuclei, notably in the system of medium mass nuclei under consideration, which is the core objective of the present work.
\begin{figure*}[!t]
	\centering
	\begin{subfigure}[t]{0.46\linewidth}
		\includegraphics[width=\linewidth]{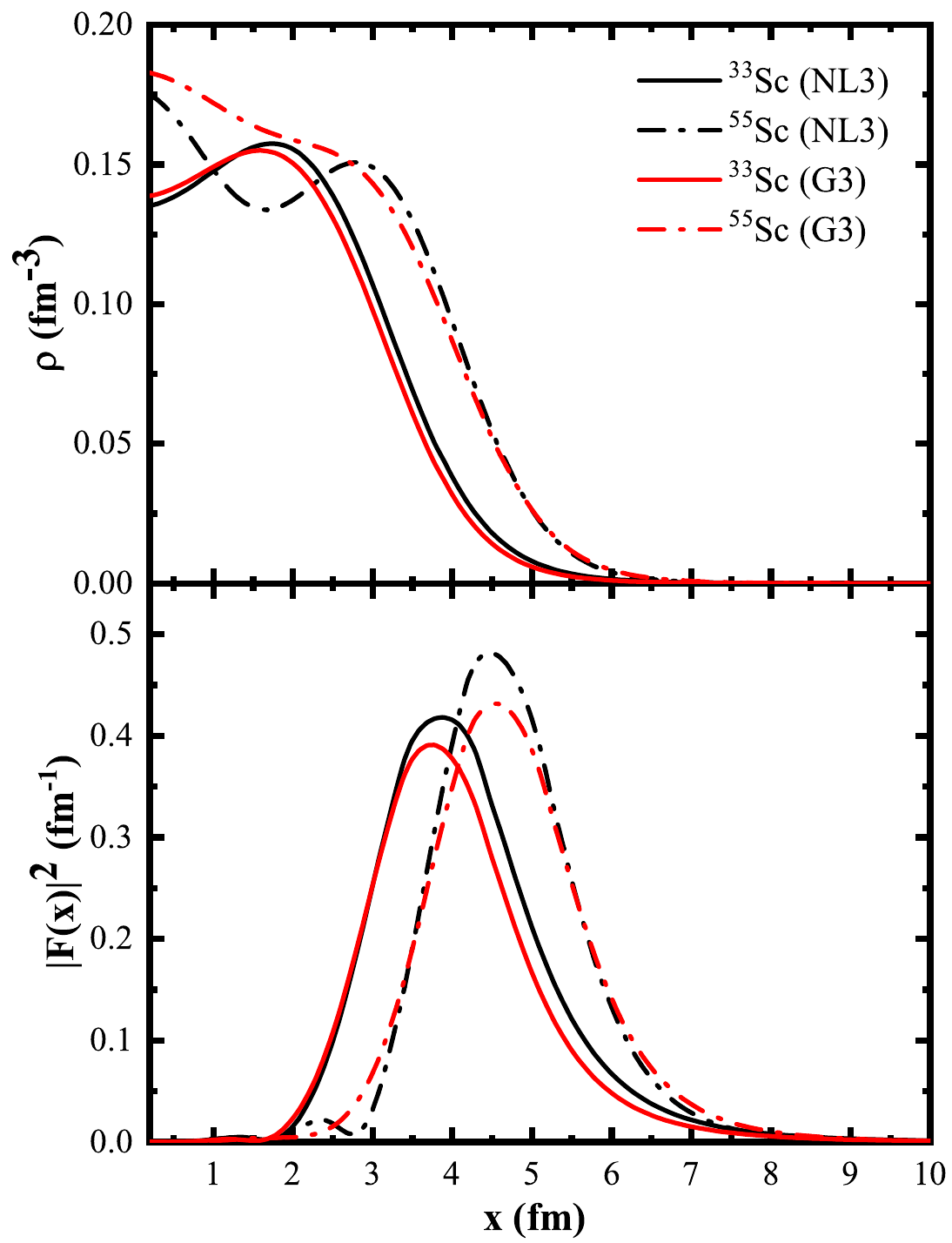}
		\caption{}
		\label{fig1a}
	\end{subfigure}
	\begin{subfigure}[t]{0.455\linewidth}
		\includegraphics[width=\linewidth]{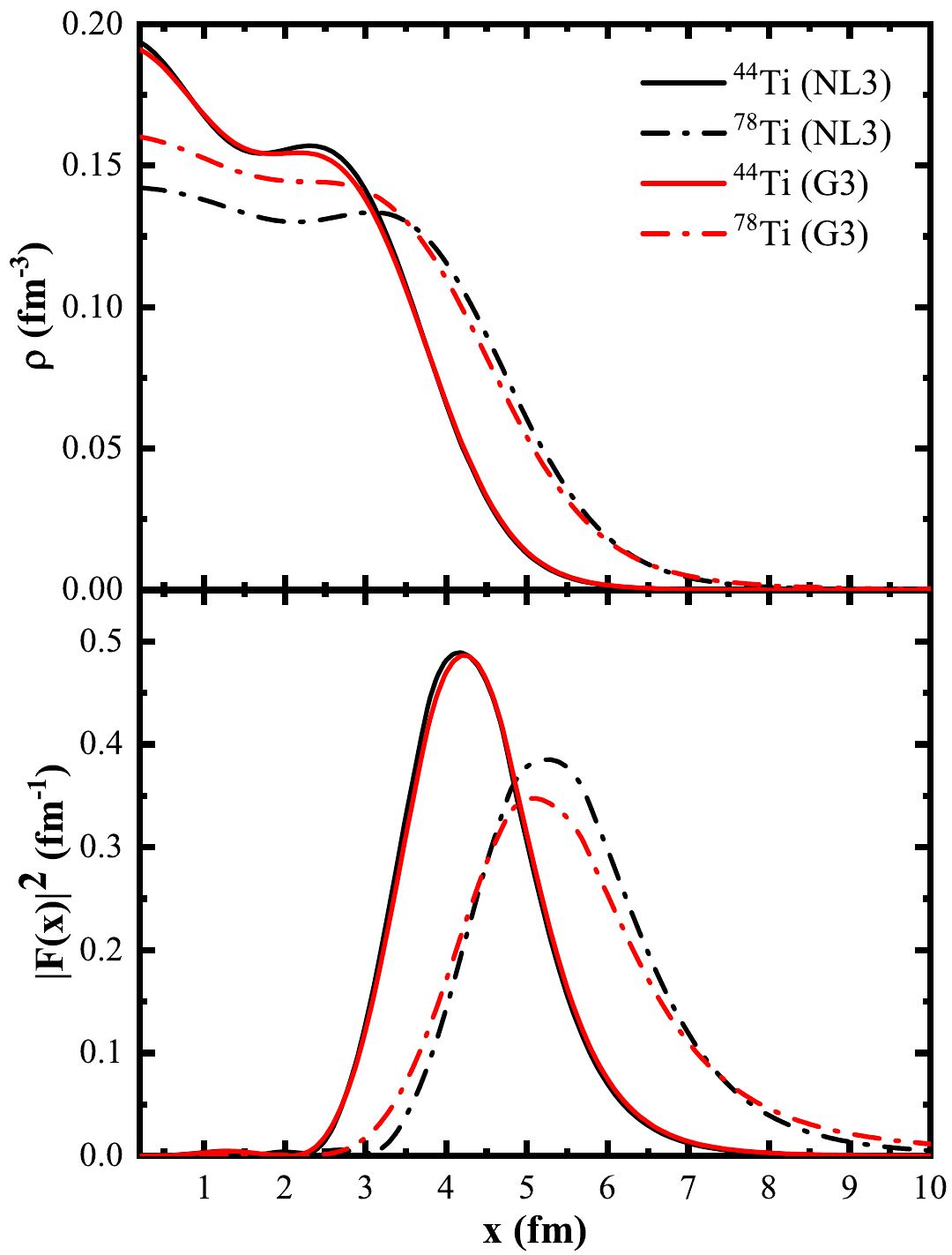}
		\caption{}
		\label{fig1b}
	\end{subfigure}
	\caption{Total density distribution (upper panel) and corresponding weight function (lower panel) as a function of nuclear distance for (a) $^{33}Sc-$ and $^{55}Sc-$ nuclei, and (b) $^{44}Ti-$ and $^{78}Ti-$ nuclei, with NL3 and G3 parameter sets. Follow the text for details.}
\end{figure*}

\section{Calculations and Results}
\label{Results} \noindent
The field equations in the effective field theory motivated relativistic mean-field (E-RMF) formalism are solved self consistently \cite{furn96,furn97,bkum17,lala97,lala09,niks02}. In a finite nucleus, the coefficient of symmetry energy is referred to as bulk property dependent on the isospin asymmetry having both surface and volume components \cite{agra12}. The volume component of symmetry energy is independent of the shape of the nucleus \cite{niko11}, inferring the lack of dependence of surface effects on volume symmetry energy. Meanwhile, the surface effect in the heavier mass regions diminishes as the surface component of symmetry energy is proportional to $A^{-1/3}$, where $A$ refers to the mass number. Thus, the presence of deformation has minimal effect on the improvement of the symmetry energy coefficient. Recently in Ref. \cite{mo15} the deformation effect of finite nuclei on the symmetry energy has been studied, which shows that with large deformation ($\beta_{2} \approx$ 0.6), the relative change in the coefficient of symmetry energy is around 0.4 MeV. Since the $\beta_{2} \ll$ 0.6 for $^{33-75}Sc$-, $^{34-80}Ti$-, $^{35-79}V$-, $^{36-80}Cr$- nuclei, we safely neglect the effects of deformation on nuclear symmetry energy. Therefore, we have considered the spherical density of the isotopes for computational ease. The isospin-dependent properties, such as symmetry energy and its derivatives, are connected with the surface properties of finite nuclear density distributions and are estimated by folding the nuclear matter properties of a given nucleus using Br\"{u}eckner's energy density functional (Br\"{u}eckner-EDF) and relativistic energy density functional (Relativistic-EDF) within the framework of CDFM.  The CDFM technique effectively addresses the variations introduced at the surface of finite nuclei by the nuclear density through the weight function \cite{gaid11,gaid12,bhuy18}.\\ \\
{\bf Nuclear density and weight function:}
We first take the density distribution obtained from the effective mean-field theory motivated relativistic mean-field (E-RMF) model. The calculated densities serve as the input to the CDFM. The next step involves the calculation of the weight function of each of the nuclei in $Sc$-, $Ti$-, $V$- and $Cr$- isotopic chains with NL3 and G3 parameter sets. Here, we use the total density ($\rho$), which is the sum of densities of neutrons ($\rho_{n}$) and protons ($\rho_{p}$). The corresponding weight function ($\vert \mathcal{F}(x) \vert^{2}$) along with the nuclear matter quantities for these nuclei are further used within the Relativistic-EDF, and Br\"{u}eckner's-EDF to calculate the symmetry energy and its components at local density. More detailed prescriptions of the CDFM formalism and calculation can be found in Refs. \cite{anto79,gaid12,gaid11,bhuy18,kaur20,qudd20,akum21,jeet22,bhuy22a}.

The upper panel of Figs. \ref{fig1a} and \ref{fig1b} presents the total density distribution ($\rho$) as a function of nuclear distance ($x$) obtained from the NL3 and G3 parameter sets of $^{33}Sc$, $^{55}Sc$ and $^{44}Ti$, $^{78}Ti$ nuclei respectively as representative case.  A careful observation of the graph reveals that with the increase in the number of protons ($Z$), minute enhancement in the surface region is observed, which is reflected in the weight function as shown in the lower panel of Figs. \ref{fig1a} and \ref{fig1b}. This enhancement means that the value of lower central density for a given nucleus reflects the lower height of the weight function. Moreover, an increase in the nuclear size results in the rightward shift of the weight function. These observations assert the importance of the total density distribution along with the weight function $\vert \mathcal{F} (x) \vert^{2}$ in the effective nuclear matter quantities. In other words, the surface density profile is imitated in the weight function, which leads to the inference of the symmetry energy and its component for a given nucleus, as discussed in subsequent paragraphs.
\begin{figure*}[!t]
	\centering
	\begin{subfigure}[t]{0.46\linewidth}
		\includegraphics[width=\linewidth]{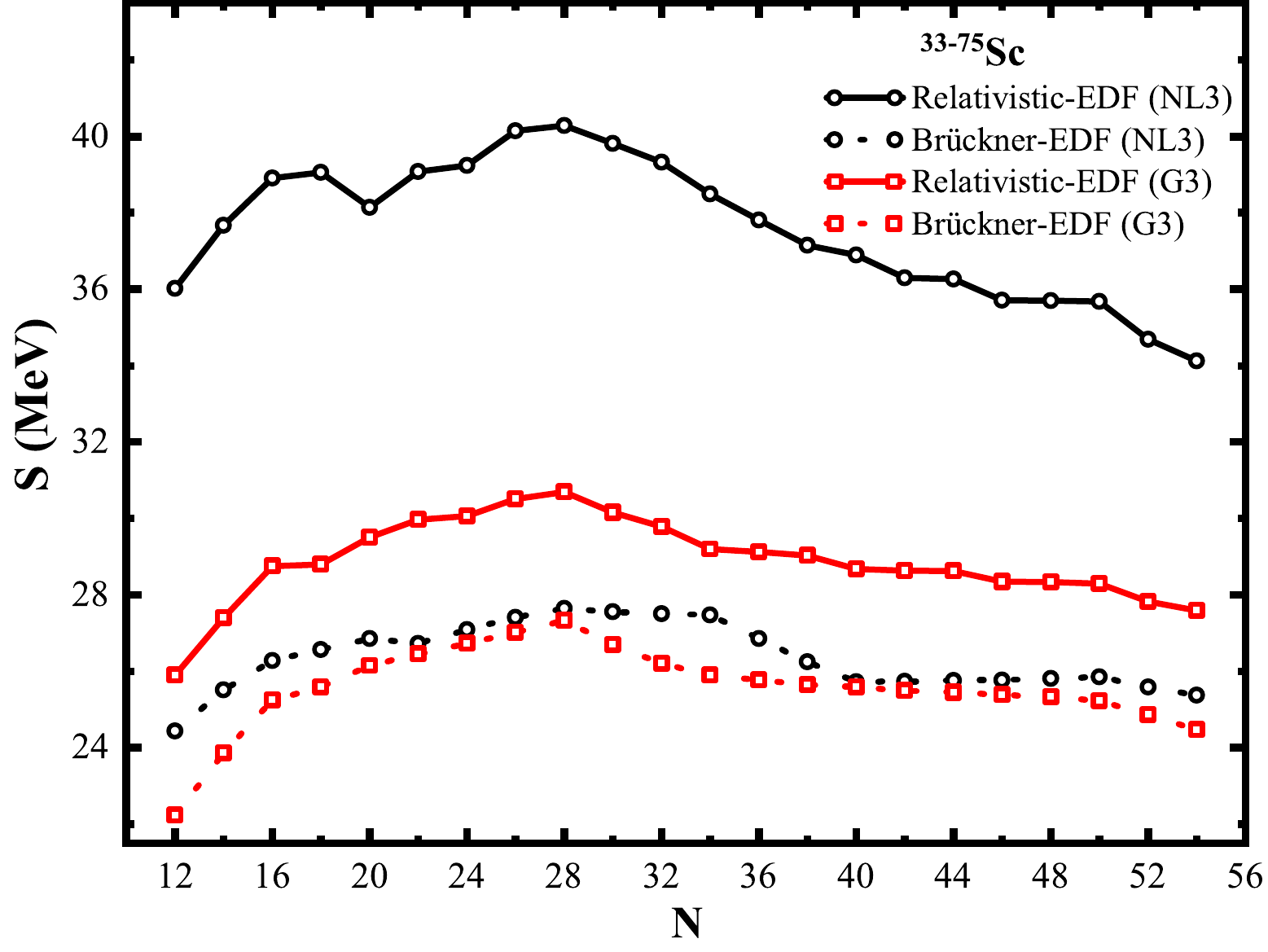}
		\caption{}
		\label{fig2a}
	\end{subfigure}
	\begin{subfigure}[t]{0.455\linewidth}
		\includegraphics[width=\linewidth]{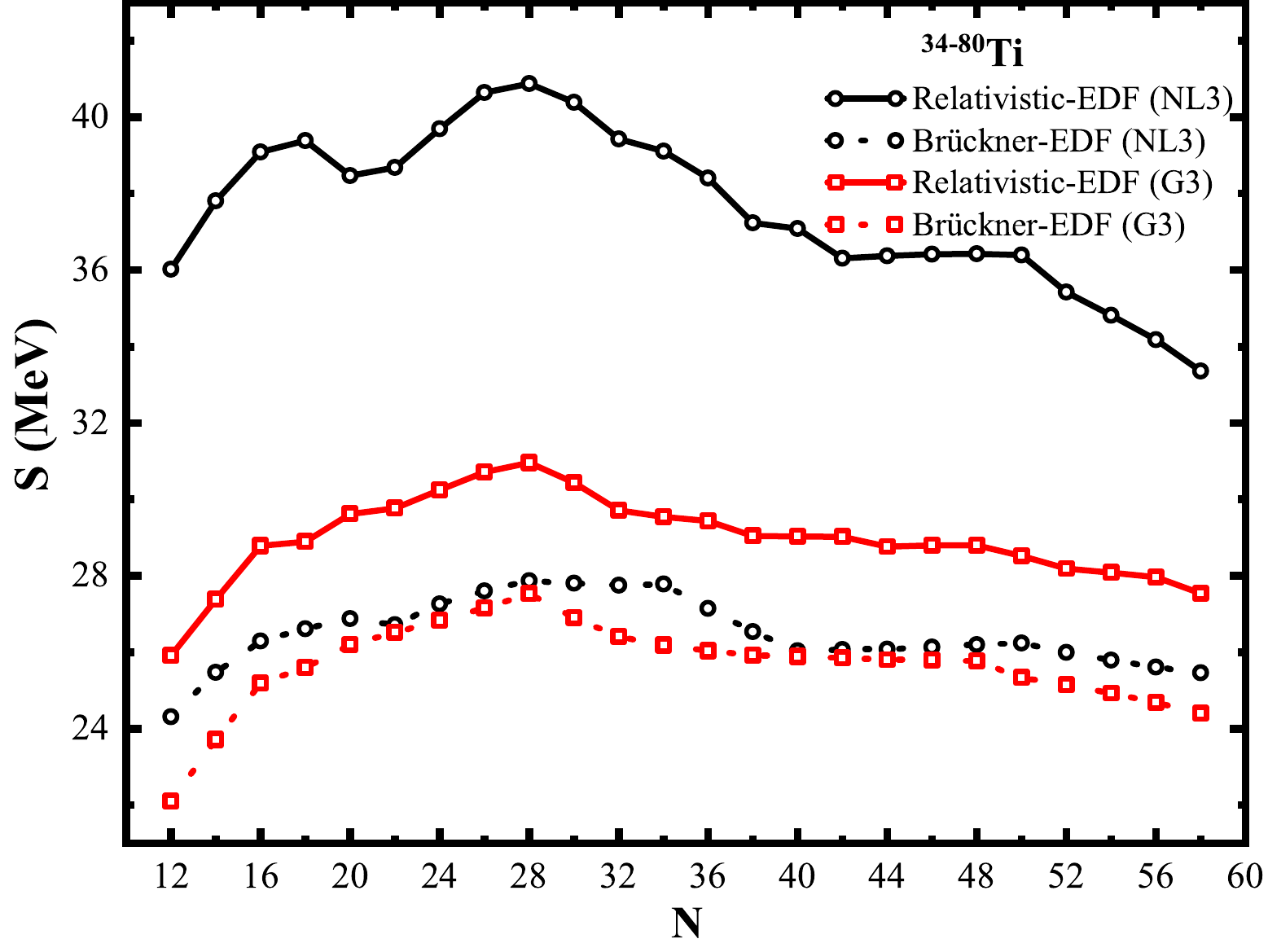}
		\caption{}
		\label{fig2b}
	\end{subfigure}
	\begin{subfigure}[t]{0.455\linewidth}
	     \includegraphics[width=\linewidth]{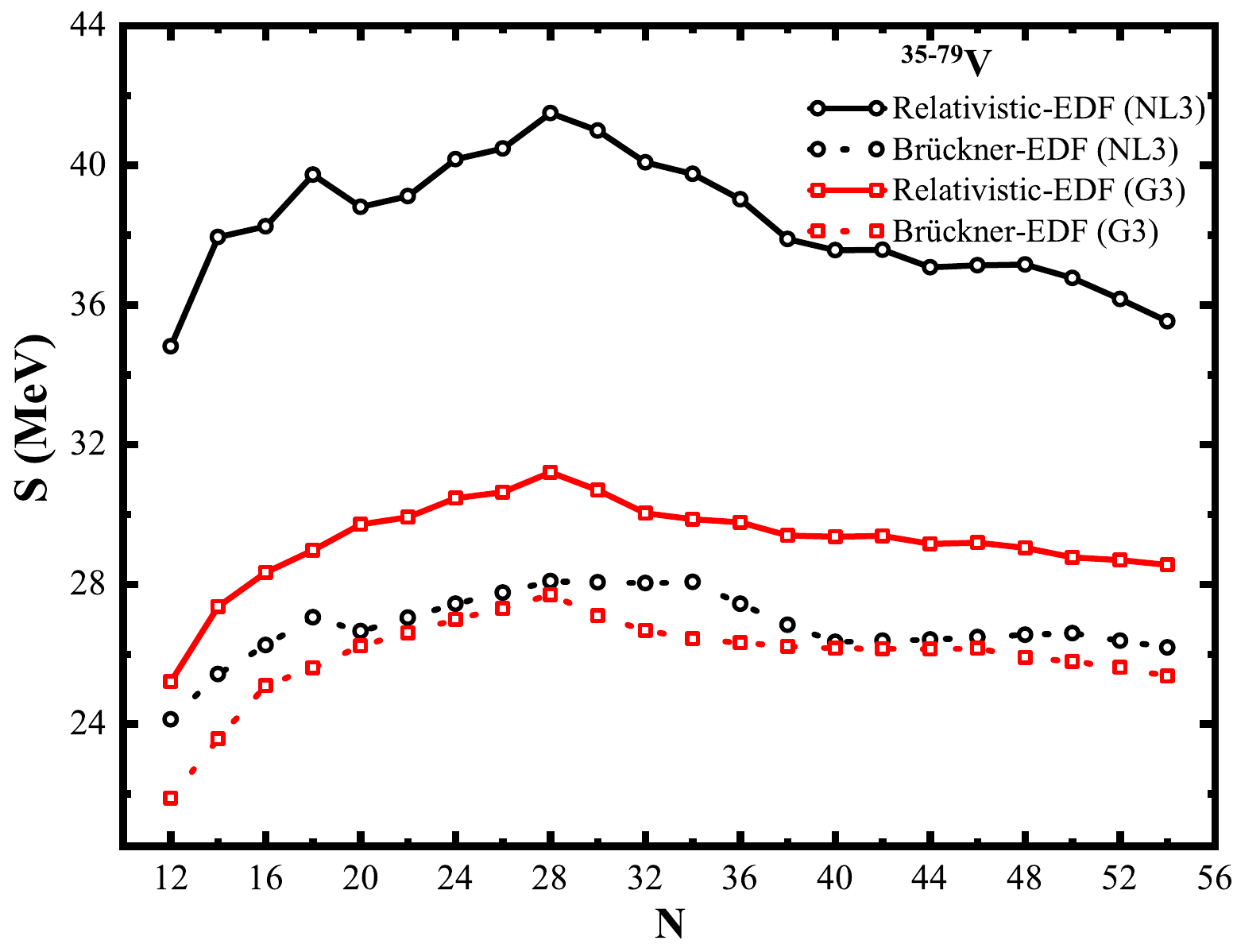}
	     \caption{}
	     \label{fig2c}
    \end{subfigure}	
	\begin{subfigure}[t]{0.455\linewidth}
	      \includegraphics[width=\linewidth]{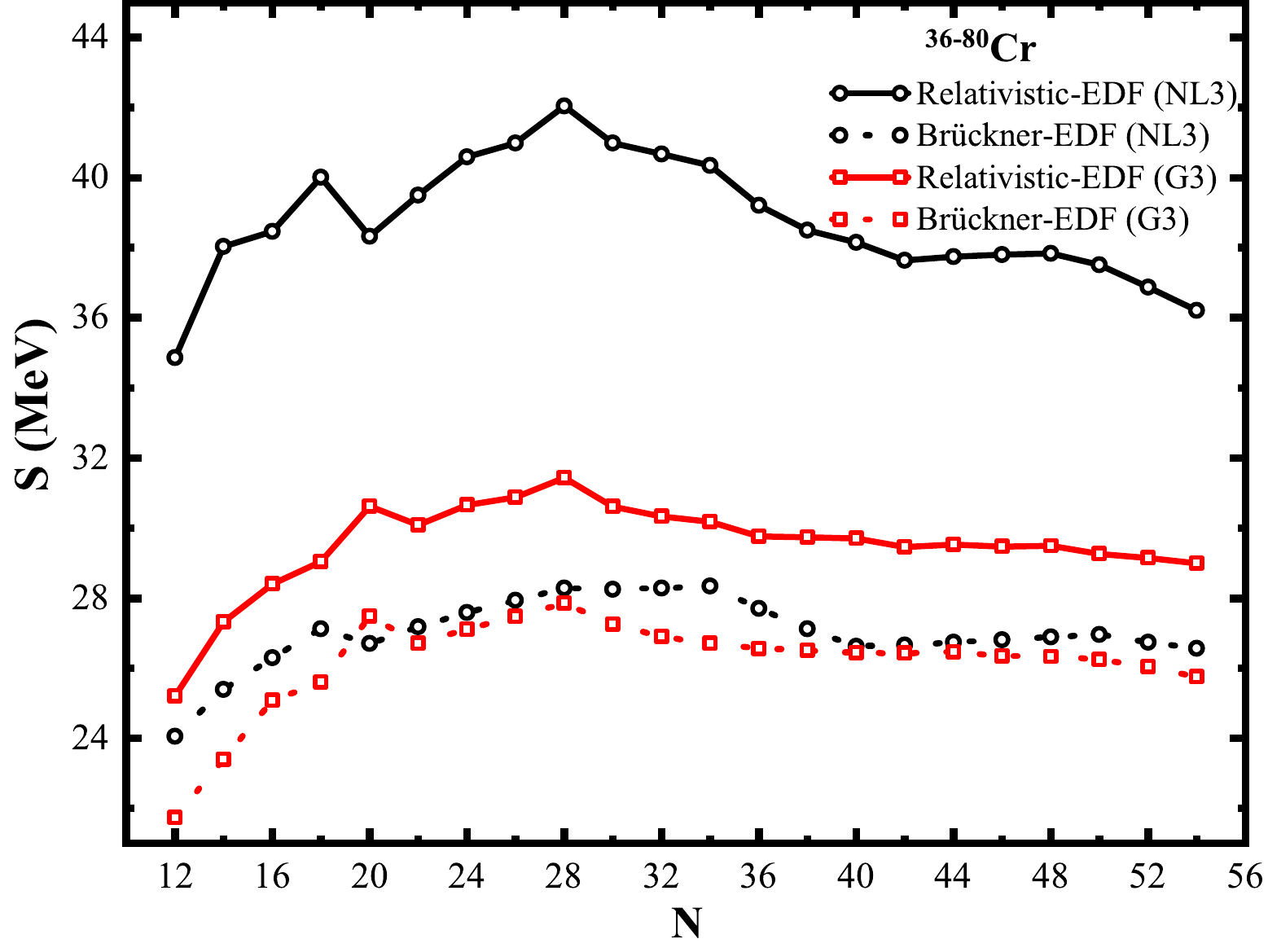}
	      \caption{}
	      \label{fig2d}
    \end{subfigure}	    
	\caption{\label{fig2} The symmetry energy $S$ is shown for (a) $^{33-75}Sc-$ isotopes ($Z$ = 21), (b) $^{34-80}Ti-$ isotopes ($Z$ = 22), (c) $^{35-79}V-$ isotopes ($Z$ = 23), and (d) $^{36-80}Cr-$ isotopes ($Z$ = 24) as a function of neutron number $N$ for the Relativistic-EDF and Br\"{u}eckner-EDF with non-linear NL3 and G3 interactions. Follow the text for details.}
\end{figure*}

\begin{figure*}[!t]
	\centering
	\begin{subfigure}[t]{0.46\linewidth}
		\includegraphics[width=\linewidth]{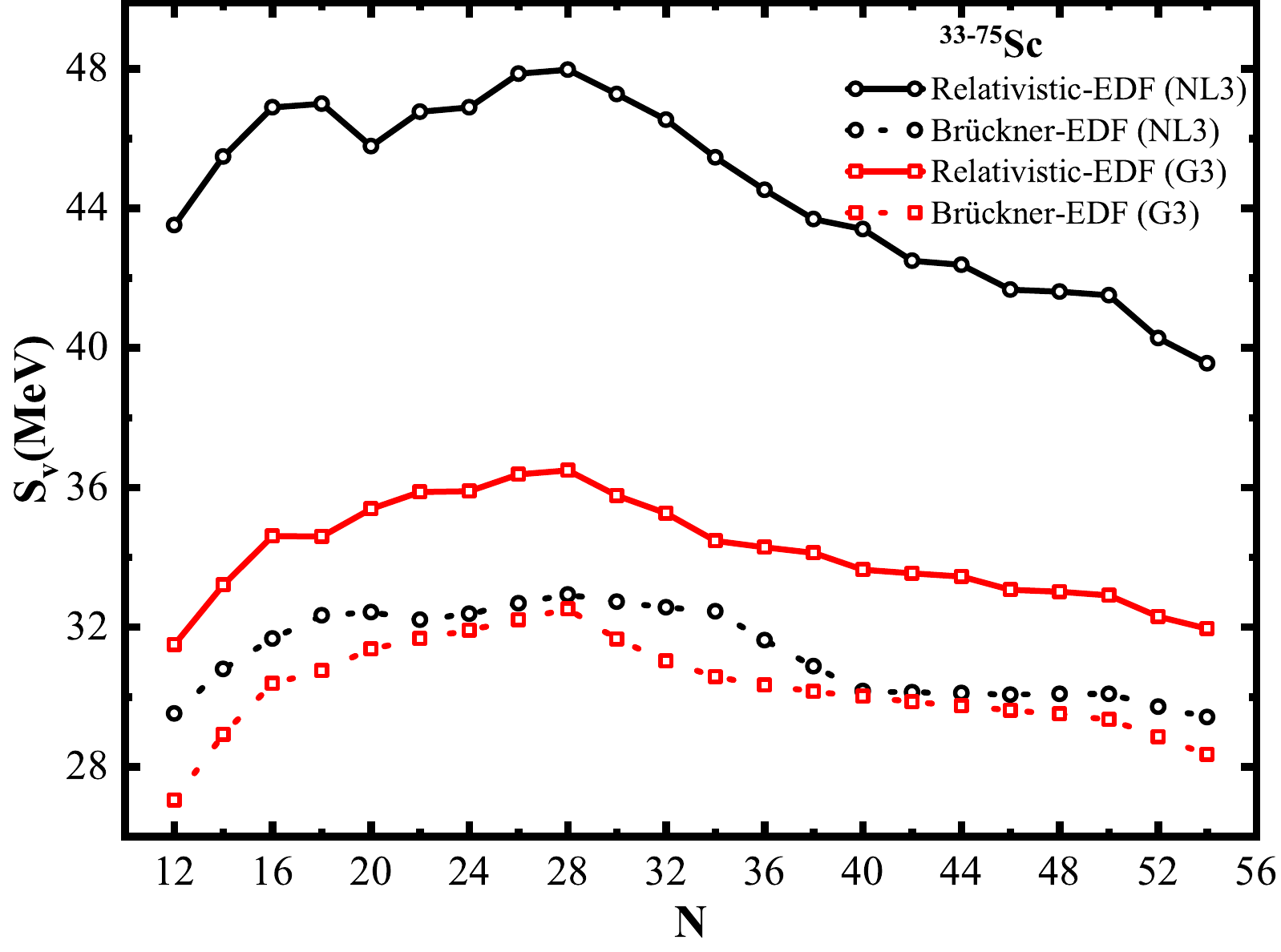}
		\caption{}
		\label{fig3a}
	\end{subfigure}
	\begin{subfigure}[t]{0.455\linewidth}
		\includegraphics[width=\linewidth]{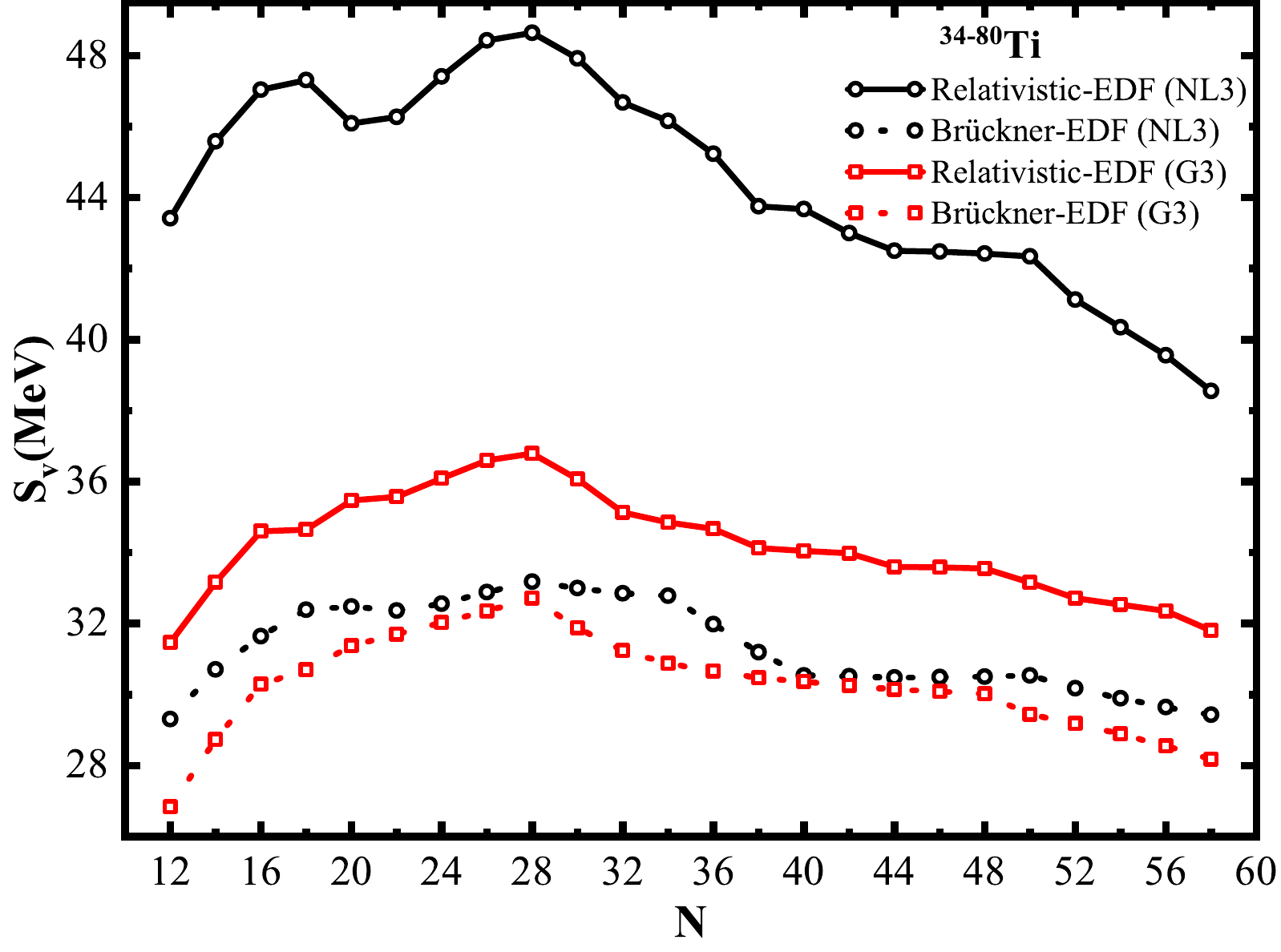}
		\caption{}
		\label{fig3b}
	\end{subfigure}
	\begin{subfigure}[t]{0.46\linewidth}
	    \includegraphics[width=\linewidth]{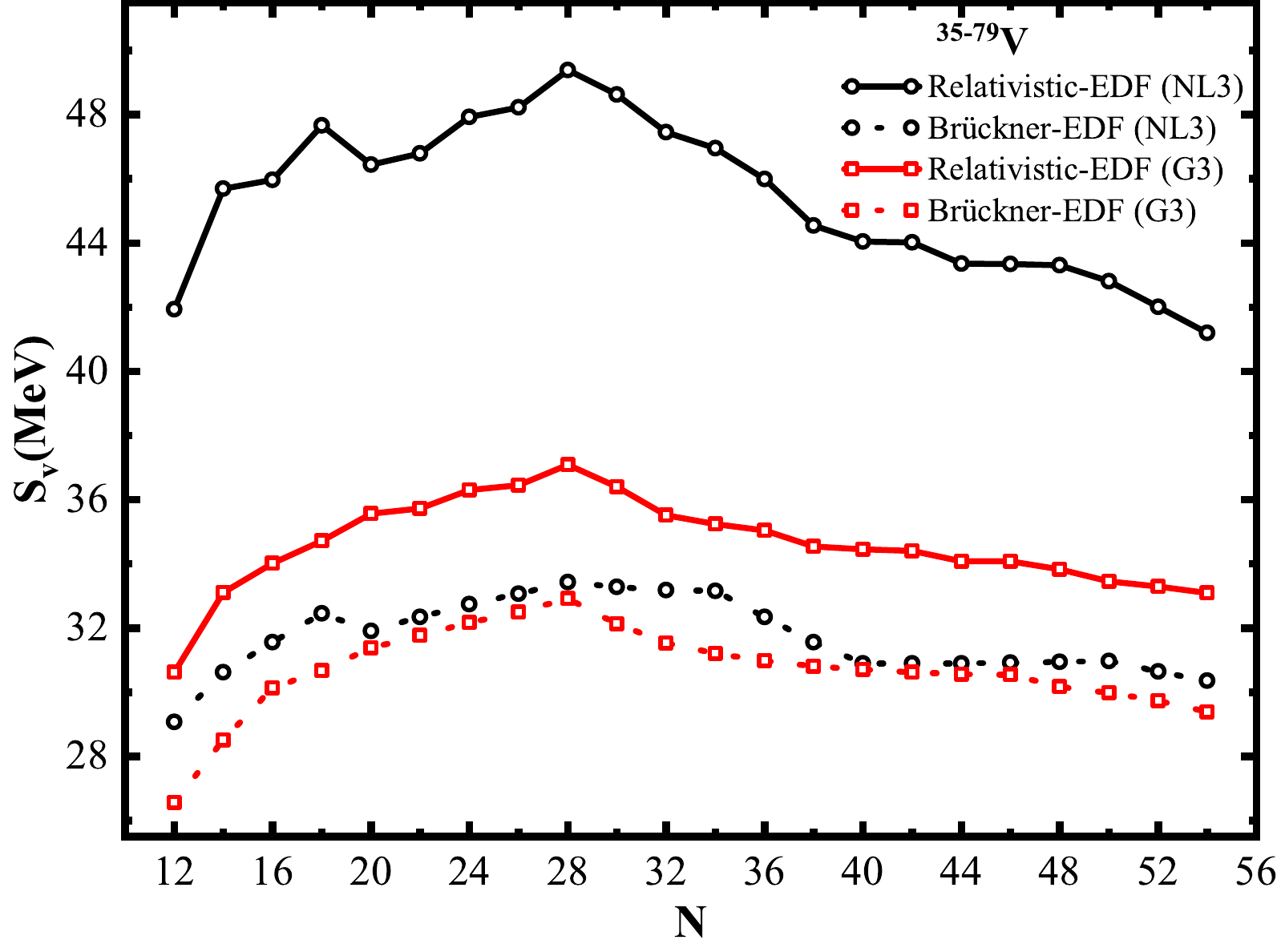}
	    \caption{}
	    \label{fig3c}
    \end{subfigure}
    \begin{subfigure}[t]{0.455\linewidth}
	    \includegraphics[width=\linewidth]{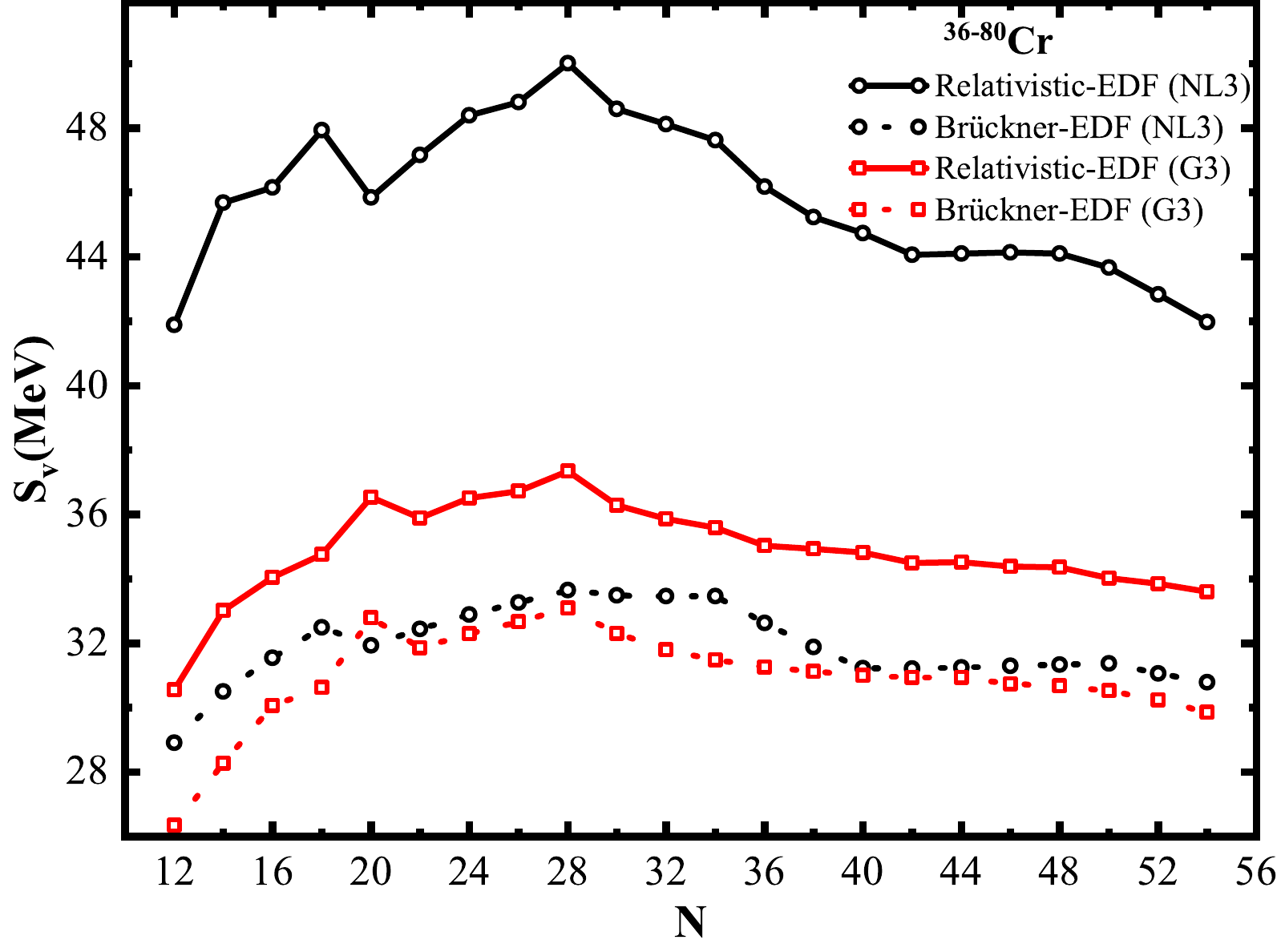}
	    \caption{}
	    \label{fig3d}
    \end{subfigure}	
	\caption{\label{fig3} The volume symmetry energy $S_{V}$ is shown for (a) $^{33-75}Sc-$ isotopes ($Z$ = 21), (b) $^{34-80}Ti-$ isotopes ($Z$ = 22), (c) $^{35-79}V-$ isotopes ($Z$ = 23), and (d) $^{36-80}Cr-$ isotopes ($Z$ = 24) as a function of neutron number $N$ for the Relativistic-EDF and Br\"{u}eckner-EDF with non-linear NL3 and G3 interactions. Follow the text for details.}
\end{figure*}

\begin{figure*}[!t]
	\centering
	\begin{subfigure}[t]{0.46\linewidth}
		\includegraphics[width=\linewidth]{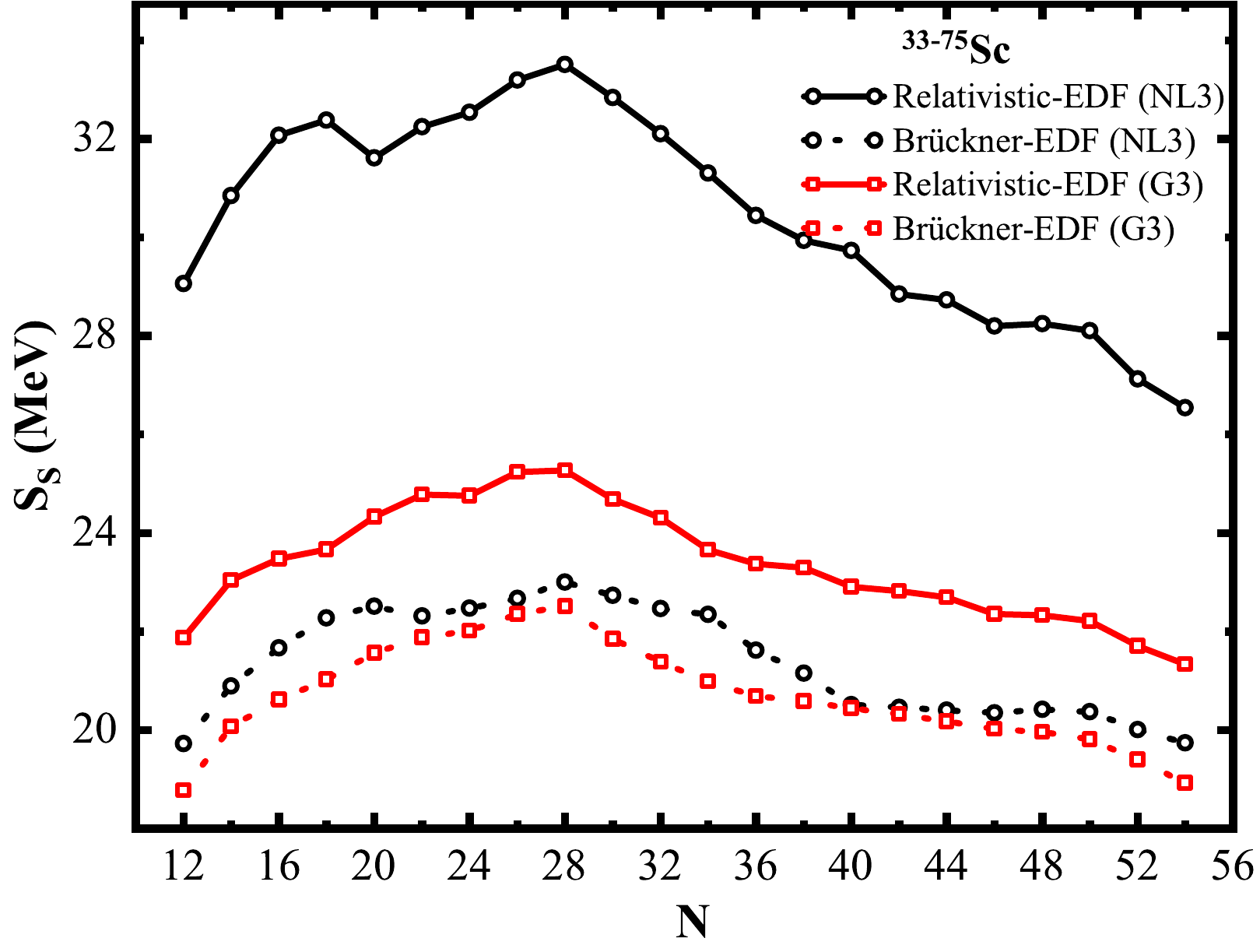}
		\caption{}
		\label{fig4a}
	\end{subfigure}
	\begin{subfigure}[t]{0.455\linewidth}
		\includegraphics[width=\linewidth]{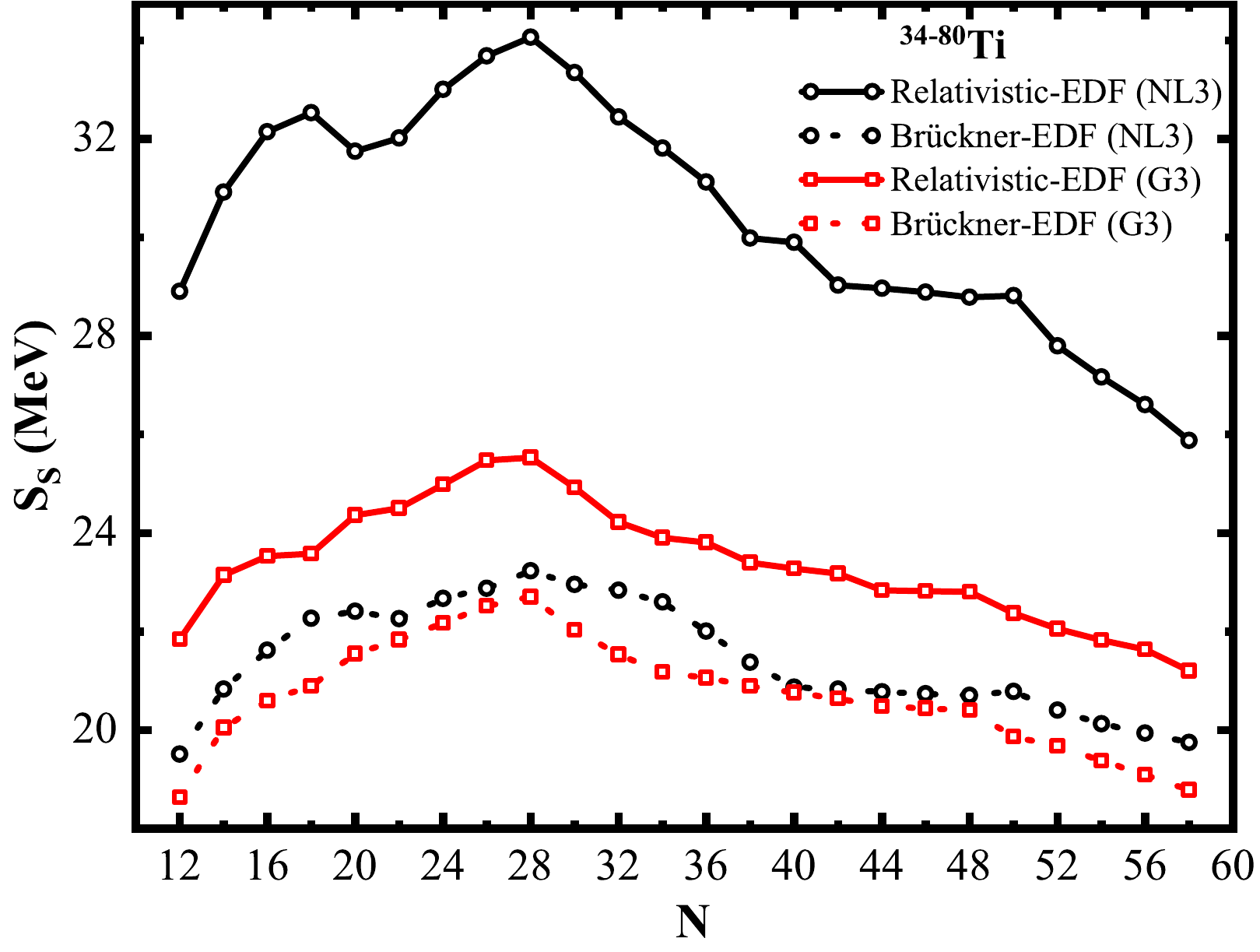}
		\caption{}
		\label{fig4b}
	\end{subfigure}
	\begin{subfigure}[t]{0.46\linewidth}
	    \includegraphics[width=\linewidth]{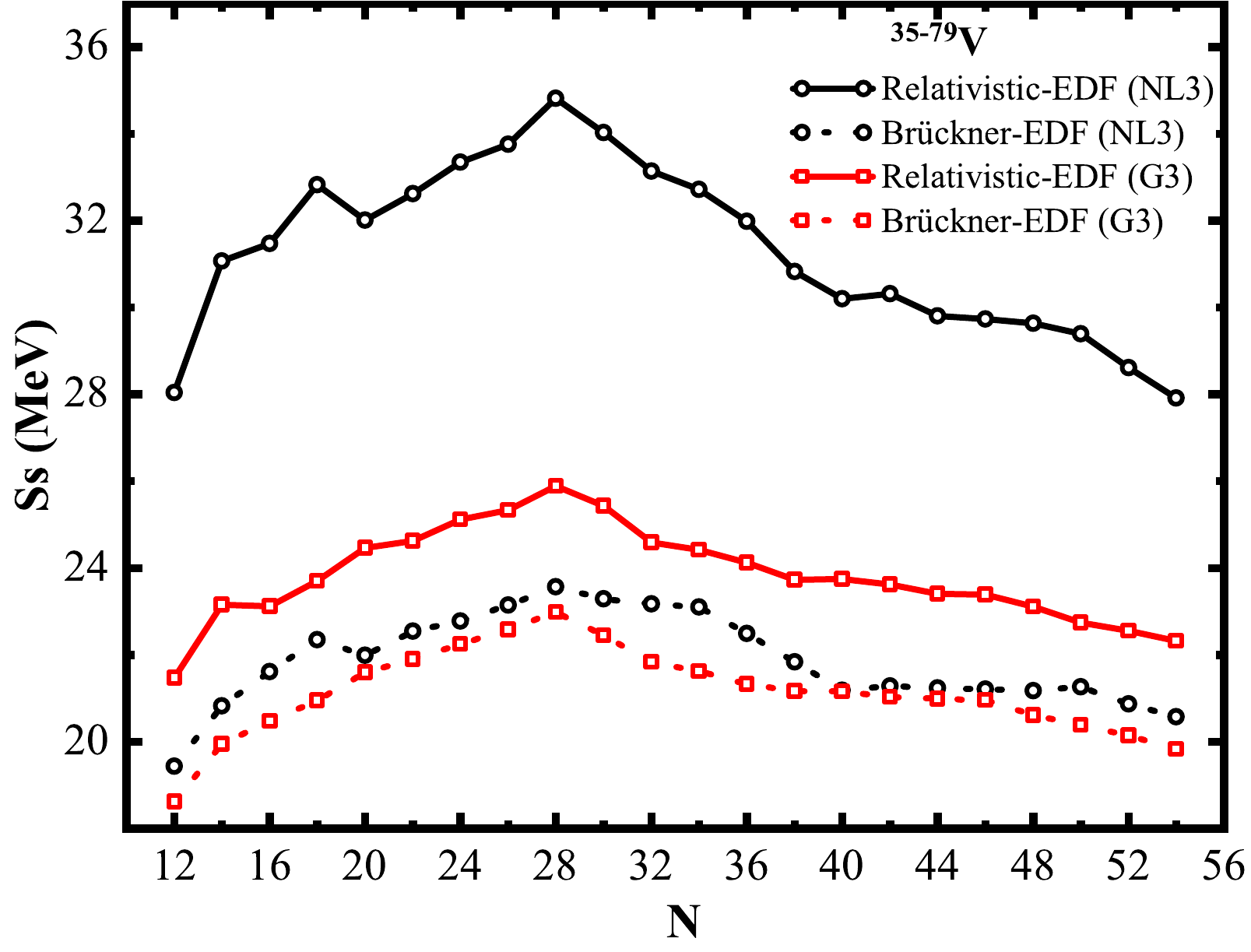}
	    \caption{}
	    \label{fig4c}
    \end{subfigure}
    \begin{subfigure}[t]{0.455\linewidth}
	    \includegraphics[width=\linewidth]{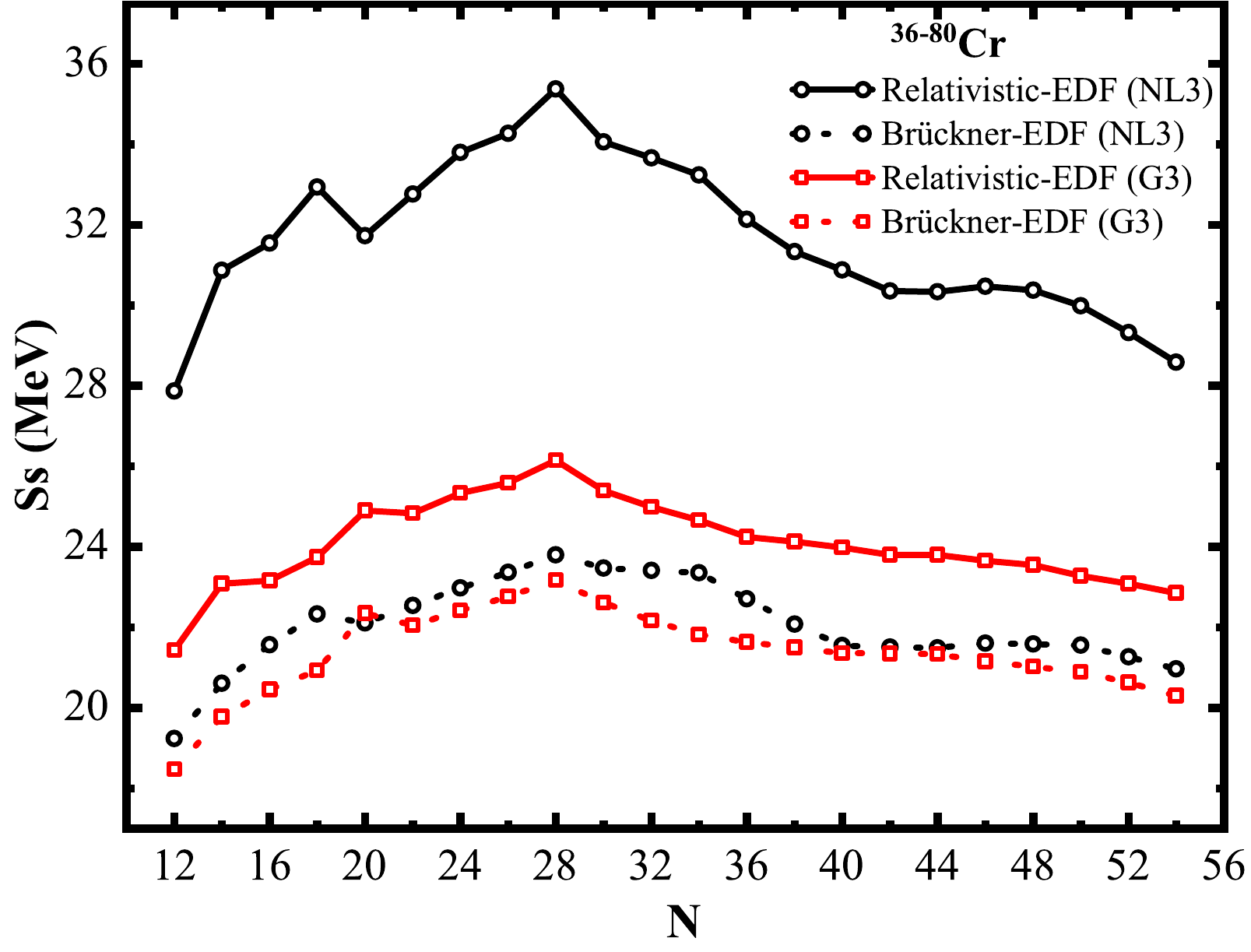}
	    \caption{}
	    \label{fig4d}
    \end{subfigure}	
	\caption{\label{fig4} The surface symmetry energy $S_{S}$ is shown for (a) $^{33-75}Sc-$ isotopes ($Z$ = 21), (b) $^{34-80}Ti-$ isotopes ($Z$ = 22), (c) $^{35-79}V-$ isotopes ($Z$ = 23), and (d) $^{36-80}Cr-$ isotopes ($Z$ = 24) as a function of neutron number $N$ for the Relativistic-EDF and Br\"{u}eckner-EDF with non-linear NL3 and G3 interactions. Follow the text for details.}
\end{figure*}

\noindent 
{\bf Symmetry energy and its components:} In recent years, the symmetry energy and its surface and volume derivatives have been successfully applied to predict the shell/sub-shell closure over the isotopic and isotonic chains for the nuclei ranging from light to super-heavy regions across the nuclear landscape \cite{kaur20,qudd20,naz19,akum21,jeet22}. In other words, the discontinuity or kink in the symmetry energy and its derivatives indicates the possible existence of shell/sub-shell closure. Here, we have adopted two formalisms, namely Br\"{u}eckner-EDF and Relativistic-EDF, through which we calculate the symmetry energy by incorporating the CDFM formalism. The aim of choosing two different approaches has been discussed in Sec. \ref{intro}. In Figs. \ref{fig2a}, \ref{fig2b}, \ref{fig2c}, \ref{fig2d} the symmetry energy ($S$) as a function of neutron number ($N$) is presented for $Sc$-, $Ti$-, $V$- and $Cr$- isotopic chains using the non-linear NL3 and G3 parameter sets. A comparison between the Br\"{u}eckner's prescription and relativistic energy density functional is made to assess the adequacy of the two prescriptions. We found that both predictions are somewhat similar, with the most distinguishing factor being the significantly wide difference in their range of values. This difference in the magnitude can be correlated with the nuclear symmetry energy of these approaches at saturation. 

For example, the CDFM results corresponding to symmetry energy using  (i) Br\"{u}eckner's prescription with the NL3 parameter set gives value as 24.4 $\leqslant S \leqslant$ 27.7 MeV for $Sc$- isotopes, 24.3 $\leqslant S \leqslant$ 27.9 MeV for $Ti$- isotopes, 24.1 $\leqslant S \leqslant$ 28.1 MeV for $V$- isotopes, and 24 $\leqslant S \leqslant$ 28.4 MeV for $Cr$- isotopes, and that of Br\"{u}eckner's prescription with G3 parameter set has values as 22.2 $\leqslant S \leqslant$ 27.4 MeV for $Sc$- isotopes, 22.1 $\leqslant S \leqslant$ 27.6 MeV for $Ti$- isotopes, 21.8 $\leqslant S \leqslant$ 27.7 MeV for $V$- isotopes, and 21.7  $\leqslant S \leqslant$ 27.9 MeV for $Cr$- isotopes and (ii) the relativistic prescription with NL3 parameter set has the values 34.1 $\leqslant S \leqslant$ 40.3 MeV for $Sc$- isotopes, 33.3 $\leqslant S \leqslant$ 40.9 MeV for $Ti$-, 34.8 $\leqslant S \leqslant$ 41.5 MeV for $V$- isotopes, and 34.8  $\leqslant S \leqslant$ 42.1 MeV for $Cr$-  isotopes and that of relativistic prescription with G3 parameter the values have the range 25.8 $\leqslant S \leqslant$ 30.7 MeV for $Sc$- isotopes, 25.9 $\leqslant S \leqslant$ 31 MeV for $Ti$- nuclei, 25.2 $\leqslant S \leqslant$ 31.2 MeV for $V$- isotopes, and 25.2  $\leqslant S \leqslant$ 31.5 MeV for $Cr$- isotopes.

Thus, we can infer that the Relativistic-EDF provides larger symmetry energy than the Br\"{u}eckner's counterpart in both the NL3 and G3 parameter sets. Moreover, in most nuclei, the G3 parameter predicts a lower value of symmetry energy than the NL3 in both the Relativistic- and Br\"{u}eckner's-EDF. We can observe from Figs. \ref{fig3} and \ref{fig4}, that sharp peaks or discontinuities are present at neutron magic number $N$ = 28 in the $Sc$-, $Ti$-, $V$- and $Cr$- isotopic chains for both the energy density functionals, signifying possible shell/sub-shell closures. More importantly, we found a significantly large kink at $N$ = 50 of $Sc$- isotopic chain with Relativistic-EDF involving NL3 and G3 parameter sets compared to Br\"{u}eckner's prescription. In the case of the $Ti$- isotopic chain, we observe a large kink at $N$ = 50 for the NL3 with Relativistic-EDF; however, the G3 parameter with Relativistic-EDF does not provide clear evidence of such discontinuity. For $V$- and $Cr$- isotopic chains we observe a minor kink near $N$ = 50 region for NL3 with Br\"{u}eckner's-EDF and Relativistic-EDF; however, the G3 parameter does not provide any evidence of discontinuity in this region. Moreover, minor discontinuity is noticed at $N$ = 40 in the relativistic prescription of $Sc$- isotopic chain for both NL3 and G3 parameter, and with $Ti$ nuclei involving NL3 parameter, which is absent in the case of Br\"{u}eckner's prescription.

Furthermore, we find minor kink at $N$ = 32 and/ or $N$ = 34 for $Ti$-, $V$- and $Cr$- isotopic chains for NL3 force parameter set for both Br\"{u}eckner's- and Relativistic- prescriptions which is in line with the experimental data \cite{rodr20,reit18,napo06,pris01,burg05}. However, no visible kinks are observed for the G3 parameter set near $N$ = 32 or $N$ = 34 along these isotopic chains. We find that the Relativistic-EDF provides marginally better results in predicting possible shell and/or sub-shell closure than the non-relativistic Br\"{u}eckner's-prescription. A similar observation was recently observed at $N$ = 126 for Pb-nuclei \cite{jeet22}. This issue is directly connected with the Coester-band problem, which is taken into account in the case of Relativistic-EDF. This work provides conclusive evidence of the superiority of the Relativistic-EDF over the other prescription.

Moreover, we have calculated the volume and surface components of symmetry energy for both relativistic and non-relativistic energy density functional, shown separately for $Sc$-, $Ti$-, $V$- and $Cr$- isotopic chains in Figs. \ref{fig3} and \ref{fig4}, respectively. From these figures, we observe that the surface and volume parameters follow similar trends with their values bound between different energies, as evident in the case of the symmetry energy (Figs. \ref{fig2a} and \ref{fig2b}). The CDFM results corresponding to volume symmetry energy $S_{V}$ and surface symmetry energy $S_{S}$ using (i) Br\"{u}eckner's prescription with NL3 parameter set gives values within the range of 29.4 $\leqslant S_{V} \leqslant $ 33 MeV and 19.7 $\leqslant S_{S}$ 23 MeV for $Sc$- isotopes, 29.3 $\leqslant S_{V} \leqslant$ 33.3 MeV and 19.5 $\leqslant S_{S} \leqslant$ 23.3 MeV for $Ti$- isotopes, 29 $\leqslant S_{V} \leqslant$ 33.5 MeV and 19.4 $\leqslant S_{S} \leqslant $ 23.6 MeV for $V$- isotopes, and 28.9 $\leqslant S_{V}  \leqslant$ 33.7 MeV and 19.2 $\leqslant S_{S} \leqslant $ 23.8 MeV for $Cr$- isotopes. The Br\"{u}eckner's prescription with G3 parameter set has values as 27 $\leqslant S_{V} \leqslant$ 32.6 MeV and 18.7 $\leqslant S_{S} \leqslant$ 22.6 MeV for $Sc$- isotopes, 26.8 $\leqslant S_{V} \leqslant$ 32.8 MeV and 18.6 $\leqslant S_{S} \leqslant$ 22.8 MeV for $Ti$- isotopes, 26.5 $\leqslant S_{V}  \leqslant$ 32.9 MeV and 18.6 $\leqslant S_{S} \leqslant $ 22.9 MeV for $V$- isotopes, and 26.3 $\leqslant S_{V}  \leqslant$ 33 MeV and 18.4 $\leqslant S_{S} \leqslant $ 23.1 MeV for $Cr$- isotopes. (ii) The relativistic prescription with NL3 parameter set has the values 39.5 $\leqslant S_{V} \leqslant$ 48 MeV and  26.5 $\leqslant S_{S} \leqslant$ 33.6 MeV for $Sc$- isotopes, 38.5 $\leqslant S_{V} \leqslant$ 48.7 MeV and 25.8 $\leqslant S_{S} \leqslant$ 34.1 for $Ti$- isotopes, and 41.1 $\leqslant S_{V}  \leqslant$ 49.3 MeV and 28 $\leqslant S_{S} \leqslant $ 34.8 MeV for $V$- isotopes, and 41.8 $\leqslant S_{V}  \leqslant$ 49.9 MeV and 27.8 $\leqslant S_{S} \leqslant $ 35.3 MeV for $Cr$- isotopes. And that of relativistic prescription with G3 parameter the values are 31.5 $\leqslant S_{V} \leqslant$ 36.5 MeV and 21.3 $\leqslant S_{S} \leqslant$ 25.3 for $Sc$- isotopes, 31.4 $\leqslant S_{V} \leqslant$ 36.8 MeV and  21.1 $\leqslant S_{S} \leqslant$ 25.6 for $Ti$- isotopes, 30.6 $\leqslant S_{V}  \leqslant$ MeV 37.1 and 21.4 $\leqslant S_{S} \leqslant $ 25.9 MeV for $V$- isotopes, and 30.5 $\leqslant S_{V}  \leqslant$ 37.3 MeV and 21.4 $\leqslant S_{S} \leqslant $ 26.2 MeV for $Cr$- isotopes. It would be interesting to have a detailed study following Ref. \cite{gaid21} regarding the use of an alternative approach in the calculation of surface and volume components of symmetry energy while using relativistic inputs in the weight function. It is expected to provide better-constrained results of the surface and volume components of symmetry energy which are consistent with the empirical data \cite{dani09}.

Similar to the trend in symmetry energy, we find that the value of volume and surface symmetry have a large difference between their relativistic and Br\"{u}eckner's prescription and that the G3 parameter shows a consistently lower value than the NL3 set. Following Ref. \cite{dani06}, volume symmetry energy ($S_V$) ranges 30 $\leq$ $S_V$ $\leq$ 32.5 and in Ref. \cite{dani09}, 31.5 $\leq$ $S_V$ $\leq$ 33.5. The calculated values of $S_V$ for G3 force parameter with Br\"{u}eckner's approach yields 26.3 $\leq$ $S_V$ $\leq$ 33 and Relativistic-EDF yields 28.9 $\leq$ $S_V$ $\leq$ 37.3, whereas NL3 force parameter with Br\"{u}eckner approach yields 28.9 $\leq$ $S_V$ $\leq$ 33.6, and with Relativistic-EDF yields 41.1 $\leq$ $S_V$ $\leq$ 49.9. We observe that the G3 parameters closely approximate other studies. However, NL3 overestimates the results which is owed to a stiffer equation of states (EoS) than G3 \cite{kuma18, lala97}.

The observation of the surface and volume symmetry energies show discontinuity or kink at $N$ = 20 and 28 for the relativistic and non-relativistic prescriptions along the $Sc$-, $Ti$-, $V$- and $Cr$- isotopic chains. We found a significantly large kink at $N$ = 50 for the $Sc$- chain involving both the NL3 and G3 parameter sets with relativistic prescription. This discontinuity at $N$ = 50 is relatively small in the case of Br\"{u}eckner's prescription. For the $Ti$- isotopic chain, we observe a large kink at $N$ = 50 for NL3 set with Relativistic-EDF; however, G3 set with Relativistic-EDF does not show clear evidence of discontinuity at $N$ = 50. Minor kink near $N$ = 50 regions is found for $V$- and $Cr$- isotopic chains for the NL3 force parameter, which is absent in G3 set. Moreover, we find a minor kink at $N$ = 40 for Relativistic-EDF of $Sc$- isotopic chain with both NL3 and G3 parameter set, and also of $Ti$- chain with NL3 set as compared to Br\"{u}eckner counterpart. Moreover, similar to the case in symmetry energy, we find a minor kink in volume and surface components of symmetry energy at $N$ = 32 and/ or $N$ = 34 for $Ti$-, $V$- and $Cr$- isotopic chains for NL3 force parameter set for both Relativistic- and Br\"{u}eckner's- prescriptions. From the theoretical analysis, the existence of possible shell and/or sub-shell closure near $N$ = 32, 34, and/ or 40 is shown in $Ti$- nuclei and $N$ = 32 and/ or $N$ = 34 for $V$- and $Cr$-  nuclei with NL3 parameter set is in accordance with the recent experimental findings \cite{mich20,rodr20,reit18,napo06,pris01,burg05}. The present calculation finds that the NL3 parameter provides a better result in reproducing the experimental data \cite{mich20} than the G3 parameter set. 

\section{Summary and Conclusions}
\label{Summary}
\noindent
In the present study, we have investigated the surface properties of $Sc$- ($Z$ = 21), $Ti$- ($Z$ = 22), $V$- ($Z$ = 23), and $Cr$- ($Z$ = 24) isotopic chains using the newly fitted Relativistic-EDF and the non-relativistic Br\"{u}eckner's energy density functional (Br\"{u}eckner-EDF). We have performed our calculations with the widely successful non-linear NL3 and recently developed G3 parameter set. We observe that the Relativistic-EDF provides a higher value of symmetry energy and its components along the $Sc$- and $Ti$- isotopic chains when compared with the Br\"{u}eckner counterpart. Moreover, the NL3 parameter shows a larger value in all the concerned nuclei in the $Sc$-, $Ti$-, $V$- and $Cr$- isotopic chains than with the G3 parameter set. The relativistic and Br\"{u}eckner's prescriptions show significant kink or discontinuity at $N$ = 20 and 28, indicating possible shell and/or sub-shell closure for the isotopic chains. Furthermore, when comparing Relativistic-EDF to its non-relativistic counterpart, a significantly larger kink at $N$ = 50 and minor evidence of discontinuity at $N$ = 40 is observed in $Sc$- nuclei with NL3 and G3 parameters, and $Ti$- nuclei with NL3 parameter set. For $V$- and $Cr$- we observe a minor kink near $N$ = 50 region for NL3 with Br\"{u}eckner's-EDF and Relativistic-EDF; however, the G3 parameter does not provide any evidence of kink in this region. However, minor kink is observed at $N$ = 32 and/ or $N$ = 34 for $Ti$-, $V$- and $Cr$- isotopic chains for NL3 force parameter set for both Relativistic- and Br\"{u}eckner's- prescriptions which are in line with the experimental findings. From the analysis, we find that the Relativistic-EDF provides a much better description of the existence of possible shell and/or sub-shell closure than the Br\"{u}eckner's prescription. This shows the superiority of Relativistic-EDF over the non-relativistic Br\"{u}eckner-EDF. The theoretical results, especially near the $N$ = 32 or 34 and 50 for these considered nuclei, can replicate the experimental findings of possible shell and/or sub-shell closure. Moreover, in the current study, the NL3 set performs slightly better at reproducing the available experimental data than the G3 parameter set. In addition, in terms of the future scope of the work, it is highly encouraged to conduct a comprehensive investigation comparing the current approach for estimating the components of symmetry energy with the newer alternative method, which aims to provide a better-constrained range for the components. This comparison firstly necessitates the incorporation of relativistic input in the weight function of the newer alternative method, which can then be extended to study various isotopic chains in different mass regions.

\section*{Acknowledgments}
This work has been supported by Science and Engineering Research Board (SERB) File No. CRG/2021/001229; FOSTECT Project Code.: FOSTECT.2019B.04; and FAPESP Project No. 2017/05660-0.

\end{document}